\documentclass[preprintnumbers,showkeys,showpacs,byrevtex,fleqn]{revtex4}
\usepackage{amsmath,amsfonts,amssymb,amscd,amsxtra,amsthm}
\usepackage{graphicx}
\usepackage{bm}
\begin{document} 
\preprint{KIAS-P12007}
\title{Heavy pseudoscalar-meson decay constants with strangeness 
\\from the extended nonlocal chiral-quark model}      
\author{Seung-il Nam}
\email[E-mail: ]{sinam@kias.re.kr}
\affiliation{School of Physics, Korea Institute for Advanced Study (KIAS), Seoul 130-722, Korea}
\date{\today}
\begin{abstract}
We study the weak-decay constants for the heavy pseudoscalar mesons, $D$, $D_s$, $B$, and $B_s$. For this purpose, we employ the extended nonlocal chiral-quark model (ExNLChQM), motivated by the heavy-quark effective field theory as well as the instanton-vacuum configuration. In addition to the heavy-quark symmetry and the nonlocal interactions between quarks and pseudoscalar mesons in ExNLChQM, a correction for the strange-quark content inside $D_s$  and $B_s$ is also taken into account and found to be crucial to reproduce the empirical values. From those numerical results, we obtain $f_{D,D_s,B,B_s}=(207.53,\,262.56,\,208.13,\,262.39)$ MeV, which are in good agreement with experimental data and other theoretical estimations. Using those numerical results, we estimate the CKM matrix elements and the Cabibbo angle with the various mesonic and leptonic heavy-meson decay channels, resulting in $(|V_{cd}|,|V_{cs}|,|V_{ub}|,|V_{td}|/|V_{ts}|)=(0.224,0.968,<5.395\times10^{-3},0.215)$ and $\theta_C=12.36^\circ$ which are well compatible with available data.
\end{abstract} 
\maketitle
\section{Introduction}
Strongly interacting systems governed by quantum chromodynamics (QCD) manifests various interesting features. In terms of the quark and gluon degrees of freedom (d.o.f.), QCD is believed to indicate the confinement as well as the asymptotic freedom. At small strong-coupling regions, perturbative treatments have been successfully applied to investigate various physics problems at high energies. In contrast, due to the nontrivial QCD vacuum structure, the low-energy quark-gluon dynamics turns out to be very  complicated, resulting in developments of lattice QCD (LQCD) simulations and effective approaches. Although LQCD has proved itself as a powerful method based on the first principle, i.e. QCD, it still contains several problematic issues, such as the notorious sign problem for finite-density QCD matters for instance. In turn, the effective approaches have been developed taking into account relevant symmetries in the nonperturbative regions and shown profound understandings for the nonperturbative QCD with many practical and successful applications~\cite{Nambu:1961fr,Nambu:1961tp,Gasser:1984gg,Gasser:1983yg}.

Among the various kinds of those effective approaches, the instanton-vacuum configuration has accumulated reliable theoretical results and interpretations for numerous physical problems, such as the spontaneous breakdown of chiral symmetry (SBCS)~\cite{Schafer:1996wv,Diakonov:1983hh,Diakonov:2002fq}, QCD vacuum properties~\cite{Dorokhov:2009zz,Dorokhov:2008zz,Nam:2008bq,Kim:2004hd}, structure functions for the mesons and baryons~\cite{Broniowski:2010nt,Dorokhov:2010zzb,Nam:2007gf,Diakonov:1996sr,Diakonov:1998ze,Ledwig:2010zq,Nam:2011yw}, QCD matter at finite temperature and/or density~\cite{Nam:2011vn,Nam:2010nk}, and so on. Note that the instanton is a semi-classical solution of the Yang-Mills equation in Euclidean space~\cite{Belavin:1975fg,'tHooft:1976fv}. The instanton vacuum is characterized by two phenomenological {\it instanton} parameters, i.e. average (anti)instanton size $\bar{\rho}\approx1/3$ fm and average inter-(anti)instanton distance $\bar{R}\approx1$ fm~\cite{Schafer:1996wv,Diakonov:1983hh,Diakonov:2002fq}. In the dilute ensemble of these pseudoparticles, representing the nontrivial QCD vacuum, the quarks are delocalized and acquire their dynamically-generated effective masses, which depend on the transferred momentum~\cite{Diakonov:2002fq}. It is worth mentioning that this momentum dependence plays the role of a UV regulator by construction so that artificial form factors for the quark-loop divergences are not necessary. 

By performing an appropriate bosonization process, we can obtain an effective chiral action (EChA), manifesting the nonlocal interactions between the quarks and pseudoscalar (PS) mesons~\cite{Diakonov:1983hh}. It has been found that EChA is very useful to study the hadron physics in terms of the PS meson and quarks. There have been many attempts to extend EChA, defined properly in Euclidean space, to Minkowski space by considering an analytic continuation between them~\cite{Praszalowicz:2001wy,Nam:2006sx}. For definiteness, we name the model defined in Minkowski space as the nonlocal chiral-quark model (NLChQM), although the naming is often used for different effective models in literatures. As shown in Ref.~\cite{Praszalowicz:2001wy,Nam:2006sx}, NLChQM has applied to the studies for the light-cone PS-meson wave function and given reliable results in comparison with other theories and experiments. Note that NLChQM inherits the specific features of the instanton model as understood, and possesses the similar symmetries and their breakdown patterns. 

Recently, much interest has been paid to the heavy meson and baryons, which contain the heavy-flavor quarks, i.e. $c$, $b$, and $t$, from theoretical~\cite{Hwang:2010hw,Rosner:2008yu,Rosner:2010ak,Choi:2007se,Badalian:2007km,Cvetic:2004qg,Wang:2005qx,Ebert:2006hj,Becirevic:1998ua,ElKhadra:1997hq,Neil:2011ku,Aoki:1997uza,Blossier:2009bx} and experimental~\cite{Abazov:2005ga,Aubert:2007bx,Ikado:2006un,Bai:1999yk,Cacciari:2005rk,GayDucati:2010dt,Korn:2003pt,Alton:2001fu} points of view. Especially, energetic studies have been done by many experimental collaborations like D$\rlap{/}{0}$~\cite{Abazov:2005ga}, BaBar~\cite{Aubert:2007bx}, Belle~\cite{Ikado:2006un}, BES~\cite{Bai:1999yk}, and so on. Also in the heavy-ion collision (RHIC)~\cite{Cacciari:2005rk} and proton-(anti)proton scattering (LHC, Tevatron)~\cite{GayDucati:2010dt,Korn:2003pt,Alton:2001fu} experiments, the heavy-hadron productions from the fire ball have been attracting interest. From the theoretical side, the nonperturbative QCD with the heavy-quark d.o.f. have been investigated actively in terms of the heavy-quark effective field theory (HQEFT)~\cite{Georgi:1990um}. In HQEFT, there appear interesting features, such as the heavy-quark symmetry and velocity super-selection rule in the heavy-quark limit $m_Q\to\infty$. The heavy-quark symmetry consists of those for the heavy quark spin and flavor. According to them, we have the mass degeneracy between the pseudoscalar (scalar) and vector (axial-vector) heavy-meson states and the heavy-quark Dirac equation without the heavy-quark current mass. Successful applications of HQEFT can be found in Refs.~\cite{Mannel:1990vg,Falk:1990pz,Maiani:1991az,Neubert:1992fk,Bagan:1992ty,Manohar:1997qy,Heitger:2003nj} and references therein.  

Considering the successes of NLChQM and HQEFT for the light- and heavy-quark sectors, respectively as mentioned above, it is quite natural and challenging to combine these two approaches and extend them to an effective model for the heavy-light quark systems. We want to name this new approach as the extended nonlocal chiral-quark model (ExNLChQM)~\cite{Nam:2011ak}. Note that there were similar tries in the contexts of the Nambu-Jona-Lasinio model~\cite{Ebert:1994tv} and effective chiral Lagrangians~\cite{Gershtein:1976mv,Khlopov:1978id}. However, ExNLChQM inherits many useful and unique benefits from its origins, i.e. NLChQM and HQEFT, such as the heavy-quark symmetry, natural UV regulator, nonlocal quark-PS meson interactions, (relatively) strong constraints on the renormalization scale, and so on, as discussed previously. We found that ExNLChQM can reproduce the weak-decay constants for the non-strange heavy PS mesons for the SU(4) $(u,d,c,b)$-flavor sector, $f_D$ and $f_B$, although there were uncertainties in the model parameters for the heavy-quark side~\cite{Nam:2011ak}. It is worth mentioning that there are alternative attempts to incorporate the heavy-light quark systems, employing the instanton physics~\cite{Musakhanov:2011xx,Musakhanov:2010zi}.

Hence, in the present work, we want to extend our previous work into the strange heavy PS mesons for $f_{D_s}$ and $f_{B_s}$ for the flavor SU(5) sector, i.e. $(u,d,s,c,b)$ flavors. It is worth mentioning that the inclusion of the strangeness even for NLChQM in the flavor SU(3) is not a simple task, according to that the effects from the $1/N_c$ corrections plays a crucial role. Thus, in order to incorporate the explicit strangeness with ExNLChQM, in principle, one needs to take into account the meson-loop corrections, corresponding to the $1/N_c$ ones, as done in Refs.~\cite{Musakhanov:2005qn,Kim:2005jc,Goeke:2007nc}. Moreover, it is worth mentioning that the axial-vector current conservation is of importance and the {\it nonlocal} contribution is necessary in order to conserve the current~\cite{Bowler:1994ir,Plant:1997jr,Nam:2007gf,Nam:2006sx}. Instead of including the meson-loop and nonlocal contributions explicitly in the present work, however, we devise a very phenomenological way to circumvent these problematic issues for ExNLChQM, using the empirical information for $f_\pi\approx132$ MeV and $f_K\approx160$ MeV and their ratios. Detailed explanations will be given in Section III.

Once having computed the theoretical results for $f_{D,D_s}$ and $f_{B,B_s}$, we can estimate the Cabibbo-Kobayashi-Maskawa (CKM) matrix elements via various leptonic and mesonic decay channels of the heavy PS mesons. Thus,  these heavy PS-meson weak-decay constants are very important physical quantities for studying $CP$ violation and have been studied extensively in numerous theoretical approaches, such as the light-cone formalism (LC)~\cite{Hwang:2010hw}, light-front quark model (LQM)~\cite{Choi:2007se}, field-correlator method (FC)~\cite{Badalian:2007km}, Bethe-Salpeter method (BS)~\cite{Cvetic:2004qg,Wang:2005qx}, relativistic quark model (RQM)~\cite{Ebert:2006hj}, QCD sum rule (QCDSR)~\cite{Penin:2001ux}, and LQCD~\cite{Becirevic:1998ua,ElKhadra:1997hq,Neil:2011ku,Aoki:1997uza,Blossier:2009bx}. From the numerical results in the present work, we obtain $f_{D,D_s,B,B_s}=(207.53,\,262.56,\,208.13,\,262.39)$ MeV, which are in good agreement with experimental and other theoretical values~\cite{Hwang:2010hw,Rosner:2008yu,Rosner:2010ak,Choi:2007se,Badalian:2007km,Cvetic:2004qg,Wang:2005qx,Ebert:2006hj,Becirevic:1998ua,ElKhadra:1997hq,Neil:2011ku,Aoki:1997uza,Blossier:2009bx}. Using the present results, we estimate the CKM matrix elements and the Cabibbo angle, resulting in $(|V_{cd}|,|V_{cs}|,|V_{ub}|,|V_{td}|/|V_{ts}|)=(0.224,0.968,<5.395\times10^{-3},0.215)$ and $\theta_C=12.36^\circ$. Again we find those numerical results are well compatible with available data~\cite{Nakamura:2010zzi}. These results indicate the reliability of the present theoretical framework. 

The present work is organized as follows: In Section II, we briefly introduce NLChQM motivated by the instanton vacuum configuration. NLChQM is extended to ExNLChQM with HQEFT in Section III. Section IV is devoted to numerical results and corresponding discussions. Summary of the present work and future perspectives are given in Section V. 
\section{Nonlocal chiral-quark model (NLChQM)}
\begin{figure}[t]
\includegraphics[width=8.5cm]{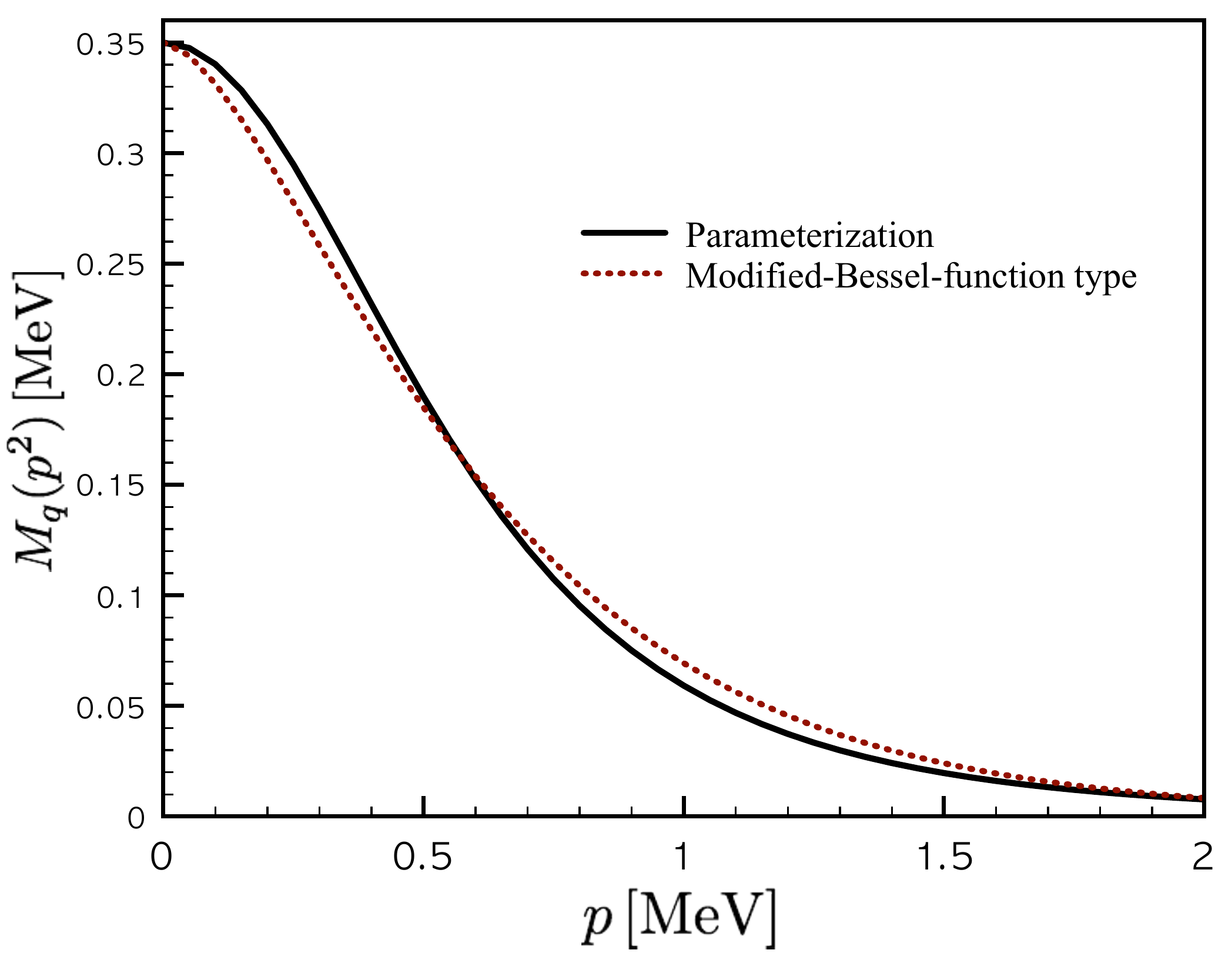}
\caption{(Color online) Effective light-quark mass $M_q$ as a function of the momentum transfer $p$, using Eq.~(\ref{eq:EFM}) (solid) and Eq.~(\ref{eq:EFMPARA}) (dot). Here, we employ $M_{q,0}=350$ MeV and $\Lambda_q=1/\bar{\rho}=600$ MeV.}       
\label{FIG0}
\end{figure}

First, we want to make a brief explanation on an effective model, motivated by the instanton vacuum configurations. The model describes the highly-nonlocal interactions between the quarks and nonperturbative gluons, i.e. the instanton solution in Euclidean space~\cite{Schafer:1996wv,Diakonov:1983hh,Diakonov:2002fq}. By a proper bosonization process similar to that for the NJL model~\cite{Nambu:1961fr,Nambu:1961tp}, integrating out all the meson fields except for the SU$(N_f)$ multiplet PS mesons, one arrives at the following effective action: 
\begin{equation}
\label{eq:ECA}
\mathcal{S}_\mathrm{eff}[\phi,m_q]=-\mathrm{Sp}\ln\left[i\rlap{/}{\partial}+i\hat{m}_q+i \sqrt{M_q}U_5\sqrt{M_q}\right],
\end{equation}
where $\phi$ and $\hat{m}_q$ stand for the light PS-meson field and current-quark mass matrix for the flavor SU(3), $\mathrm{diag}(m_u,m_d,m_s)$. $\mathrm{Sp}$ and $M_q=M_q(\partial^2)$ denote the functional trace over relevant spin indices and momentum-dependent effective-quark mass for the light-flavor quarks $q=(u,d,s)$, respectively. Generically, $M_q$ is defined by the modified Bessel functions in momentum space by~\cite{Diakonov:2002fq}
\begin{equation}
\label{eq:EFM}
M_q(p^2)=M_{q,0}F^2(p^2)=
2t\left[I_0(t)K_1(t)-I_1(t)K_0(t)-\frac{1}{t}I_1(t)K_1(t) \right],
\,\,\,\,t=\frac{\sqrt{p^2}\bar{\rho}}{2}.
\end{equation}
We note that $F(p^2)$ comes from the Fourier transform of the instanton zero mode~\cite{Diakonov:2002fq}. In many applications, $M_q$ is parameterized for simplicity as follows:
\begin{equation}
\label{eq:EFMPARA}
M_q(p^2)=M_{q,0}\left(\frac{2\Lambda^2_q}{2\Lambda^2_q+p^2} \right)^2.
\end{equation}
In the present work, we will use Eq.~(\ref{eq:EFMPARA}) for numerical calculations, since there are only small differences between the usages of Eqs.~(\ref{eq:EFM}) and (\ref{eq:EFMPARA}) as shown in Figure~\ref{FIG0}. The self-consistent equation of the model for the chiral limit reads~\cite{Diakonov:2002fq}:
\begin{equation}
\label{eq:SDL}
\frac{1}{\bar{R}^4}=4N_c\int_E\frac{d^4p}{(2\pi)^4}\frac{M^2_q}{p^2+M^2_q}.
\end{equation}
If we make use of the phenomenological instanton parameters $\bar{R}\approx1$ fm and $\bar{\rho}\approx1/3$ fm, the value of $M_{q,0}$ is determined to be about $350$ MeV via Eq.~(\ref{eq:SDL}). Thus, we will use this value for $M_{q,0}$ and a renormalization  scale of the model $\Lambda_q\approx1/\bar{\rho}\approx600$ MeV throughout the present work. 

Taking into account the nonzero current-quark mass beyond the chiral limit, the weak-decay constant for the PS meson, we have the following formula for the leading {\it local} contribution for $f_\pi$ and $f_K$:
\begin{equation}
\label{eq:FPI}
f^2_{\phi}=4N_c\int_E\frac{d^4p}{(2\pi)^4}
\frac{\mathcal{N}_f\mathcal{N}_g\sqrt{M_fM_g}
\left[\bar{M}_f+\bar{M}_g-\frac{|p|}{4}(M'_f+M'_g) \right]}
{(p^2+\bar{M}^2_f)(p^2+\bar{M}^2_g)},
\end{equation}
where we have defined the notations as $\bar{M}_q=M_q+m_q$ and $M'=\partial M/\partial |p|$. The subscripts $f$ and $g$ denote the quark for each flavor inside the PS meson $\phi\sim f\bar{g}$. Throughout this work, we employ the normalization as $f_\pi=\sqrt{2}F_\pi$, where $F_\pi\approx93$ MeV. Note that we introduced a phenomenological multiplicable factor $\mathcal{N}_q$ and will explain the meaning of this parameter in detail below. First, choosing $\mathcal{N}_f=\mathcal{N}_g=1$, we have the following numerical values using Eq.~(\ref{eq:FPI}):
\begin{equation}
\label{eq:FPIK}
f_\pi\approx f_K\approx113\,\mathrm{MeV}
\,\,\,\,\mathrm{for}\,\,\,\,m_{u,d,s}=(5,5,100)\,\mathrm{MeV}.
\end{equation}
In order to obtain the empirical value for $f_\pi\approx132$ MeV, one has $\mathcal{N}_{u,d}=1.17$, taking into account the isospin symmetry $m_u=m_d$ as in Eq.~(\ref{eq:FPIK}). As for the kaon, $\mathcal{N}_s=1.71$ reproduces the empirical value for $f_K\approx160$ MeV. 
Below,  we want to explain the physical meanings of these $\mathcal{N}_f$ parameters  in detail:
\begin{itemize}
\item In order for satisfying the axial-vector-current conservation for the PCAC relation in the nonlocal effective model in terms of the flavor SU(3) symmetry, one needs to include {\it nonlocal} contributions additionally to Eq.~(\ref{eq:FPI})~\cite{Bowler:1994ir,Plant:1997jr,Nam:2007gf,Nam:2006sx}, although we have ignored them for simplicity. They provide about $(10\sim30)\,\%$ increase of relevant physical quantities in this model.  Hence, $\mathcal{N}_{u,d}$ slightly bigger than unity compensate the absence of the nonlocal contributions in Eq.~(\ref{eq:FPI}). This simplification can be understood easily by the following:
\begin{equation}
\label{eq:PARA1}
f_\phi=f^{\mathrm{L}}_\Phi+f^{\mathrm{NL}}_\phi
\to\mathcal{N}_\alpha\mathcal{N}_\beta f^{\mathrm{L}}_\phi,
\end{equation}
where the superscripts L and NL indicate the leading local and sub-leading nonlocal contributions, respectively, while the subscripts $\alpha$ and $\beta$ represent the quark flavors inside the PS meson in the flavor SU(3) symmetry $m_u\sim m_d\sim m_s\ll\Lambda_q$.
\item If we go beyond the flavor SU(3) symmetry and the heavier strange quark comes into play, breaking the symmetry $(m_{u,d}\ll m_s)$, the situation gets complicated: One needs to consider the $1/N_c$ corrections~\cite{Goeke:2007nc,Nam:2008bq}, which also corresponds to the meson-loop corrections (MLC). Therefore, $\mathcal{N}_s$, which is considerably larger than unity, plays the role of MLC effectively as
\begin{equation}
\label{eq:PARA2}
f_{\phi_s}\approx\mathcal{N}_\alpha\mathcal{N}_\beta f^{\mathrm{L}}_{\phi_s}+f^{\mathrm{MLC}}_{\phi_s}\to \mathcal{N}_\alpha\mathcal{N}_sf^{\mathrm{L}}_{\phi_s}
\,\,(\mathrm{or}\,\,\mathcal{N}_\beta\mathcal{N}_sf^{\mathrm{L}}_{\phi_s}).
\end{equation}
Here, the subscript $\phi_s$ stands for the PS meson including the heavier strange quark, i.e. the kaon.
\end{itemize}
Thus, $\mathcal{N}_q$ can be understood as a very phenomenological compensation for the above two simplifications to reproduce the empirical data. The differences in $\mathcal{N}_{u,s,d}$ may also indicate the distinctive interaction strengths between the flavors and the QCD vacuum, as observed in usual constituent-quark models. In what follows, $\mathcal{N}_{u,s,d}$ will be used for computing the heavy-meson decay constants with strangeness, instead of cosidering the nonlocal ($F^\mathrm{L}_\Phi$) and MLC ($F^\mathrm{MLC}_\Phi$) contributions explicitly. We verified that this phenomenological approach makes the problems much simpler to a great extent analytically as well as numerically. Moreover, the numerical results obtained from this approach are turned out to be qualitatively-well compatible with other experimental and theoretical studies as shown in Section IV. 
\begin{table}[h]
\begin{tabular}{c|c|c|c|c|c|c|c|c}
$\hspace{0.25cm}m_u\hspace{0.25cm}$
&$\hspace{0.25cm}m_d\hspace{0.25cm}$
&$\hspace{0.25cm}m_s\hspace{0.25cm}$
&$\hspace{0.25cm}m_c\hspace{0.25cm}$
&$\hspace{0.25cm}m_b\hspace{0.25cm}$
&$\hspace{0.25cm}M_D\hspace{0.25cm}$
&$\hspace{0.25cm}M_{D_s}\hspace{0.25cm}$
&$\hspace{0.25cm}M_B\hspace{0.25cm}$
&$\hspace{0.25cm}M_{B_s}\hspace{0.25cm}$\\
\hline
$5$&$5$&$100$&$1270$&$4670$&$1867$&$1968$&$5279$&$5366$\\
\end{tabular}
\caption{Numerical inputs for the relevant masses [MeV] for the calculations~\cite{Nakamura:2010zzi}.}
\label{TABLE1}
\end{table}

\section{Extended nonlocal chiral quark model}
Now, we are in a position to introduce ExNLChQM as derived in the previous work~\cite{Nam:2011ak}. We note that this model is in principle equivalent to that suggested in Ref.~\cite{Ebert:1994tv} in many aspects, besides several specific features, i.e. quark-PS meson nonlocal interactions and natural UV regulator by construction for instance. Those features are inherited from the instanton-vacuum effects. By taking into account the $(u,d,s,c,b)$ flavors and the structure of Eq.~(\ref{eq:ECA}), one can construct the following effective chiral Lagrangian for the heavy $(Q)$ and light $(q)$ quark system in Minkowski space:
\begin{equation}
\label{eq:LAG}
\mathcal{L}^\mathrm{ExNLChQM}_{\mathrm{eff}}[\Phi,m]=\bar{\psi}
\left[i\rlap{\,/}{D}-\hat{m}_q-\hat{m}_Q-\sqrt{\mathcal{M}}^\dagger\mathcal{U}_5\sqrt{\mathcal{M}} \right]\psi,
\end{equation}
where $\Phi$ stands for the heavy ($H$) and light ($L$) mesons. $\hat{m}_{q,Q}$ indicate the current-quark mass matrices for the light- and heavy-flavor quarks for $q=(u,d,s)$ and $Q=(c,b)$: $\hat{m}_q=\mathrm{diag}(m_u,m_d,m_s,0,0)$ and $\hat{m}_q=\mathrm{diag}(0,0,0,m_c,m_b)$. The quark spinor is assigned as $\psi=(u,d,s,c,b)^T$. Although the effective Lagrangian in Eq.~(\ref{eq:LAG}) has a SU(5) symmetric form for the flavors, the symmetry is broken explicitly by the difference between the current-quark masses as in Table~\ref{TABLE1}. Moreover, the flavor SU(5) symmetry is broken further by the distinctive effective quark masses for the heavy and light quarks as will be discussed later. Note that the chiral symmetry is explicitly broken as understood by Eq.~(\ref{eq:LAG}) and SBCS is emerged by finite $\mathcal{M}$ values, corresponding to the effective quark mass, $\mathcal{M}\equiv\mathrm{diag}(M_q,M_Q)=\mathrm{diag}(M_u,M_d,M_s,M_c,M_b)$ as in Eq.~(\ref{eq:EFM})~\cite{Nam:2011ak}. We define the effective heavy-quark mass, assuming it possesses the same analytic structure with that for the light quark, in Minkowski space:
\begin{equation}
\label{eq:MMMH}
M_Q(p^2)=M_{Q,0}\left(\frac{2\Lambda^2_Q}{2\Lambda^2_Q-p^2} \right)^2,
\end{equation}
where $M_{Q,0}$ indicates the effective heavy-quark mass at zero virtuality. $\Lambda_Q$ denotes a renormalization scale for the heavy quarks. The nonlinear heavy-light PS-meson fields for the SU(5) symmetry can be constructed as follows:
\begin{equation}
\label{eq:U5}
\mathcal{U}_5=\exp\left[\frac{i\gamma_5\mathrm{U}_5}{F_{\Phi}}\right].
\end{equation}
Here, the weak-decay constant is assigned by the normalization $F_{\Phi}\equiv f_\Phi/\sqrt{2}$ as mentioned. The explicit form of the $24$-plet $\mathrm{U}_5$ matrix is given in Appendix. This sort of the extension of a flavor group is in principle equivalent to that suggested in Ref.~\cite{Gamermann:2007fi}.

Since we are interested in computing the weak-decay constants in the present work, we expand the nonlinear PS-meson fields in the effective Lagrangian up to $\mathcal{O}(\Phi^1)$ for the current $\propto\langle0|J_W|\Phi\rangle$. Then, the effective Lagrangian in Eq.~(\ref{eq:LAG}) can be represented in three separate parts, i.e. light-light (LL), heavy-heavy (HH), and heavy-light (HL, LH) quark terms as done in Ref.~\cite{Nam:2011ak}:
\begin{eqnarray}
\label{eq:EFL1}
\mathcal{L}_\mathrm{eff}^\mathrm{ExNLChQM}
&=&\mathcal{L}^\mathrm{LL}_\mathrm{eff}
+\mathcal{L}^\mathrm{HH}_\mathrm{eff}+\mathcal{L}^\mathrm{(HL,LH)}_\mathrm{eff}
\cr
&=&
\left[\bar{q}\left(i\rlap{/}{\partial}_q-m_q-M_q \right)q
-\frac{1}{F_L}\bar{q}\sqrt{M_q}
\left[i\gamma_5L \right]
\sqrt{M_q}q \right]_\mathrm{LL}
+\left[\bar{Q}\left(i\rlap{/}{\partial}_Q-m_Q-M_Q \right)Q\right]_\mathrm{HH}
\cr
&-&\left[\frac{1}{F_H}\bar{Q}\sqrt{M_Q}
\left[i\gamma_5H
 \right]
\sqrt{M_q}q\right]_\mathrm{HL}
-\left[\frac{1}{F_H}\bar{q}\sqrt{M_q}
\left[i\bar{H}\gamma_5 \right]
\sqrt{M_Q}Q\right]_\mathrm{LH},
\end{eqnarray}
where $L$ and $H$ denote the light and heavy PS-meson fields defined in Eq.~(\ref{eq:U5F}) in Appendix. Note that, however, we have not considered the $Q\bar{Q}$-meson states, corresponding to the quarkonia states, in the above effective Lagrangian. In Eq.~(\ref{eq:EFL1}), we have also taken into account the effective heavy-quark mass term, $M_Q$ in $\mathcal{L}^\mathrm{HH}_\mathrm{eff}$, in addition to the current mass $m_Q$. According to HQEFT, one can redefine the heavy-quark and heavy-meson fields as follows~\cite{Ebert:1994tv}:
\begin{equation}
\label{eq:HMFD}
Q(x)=\frac{1+\rlap{/}{v}}{2}e^{-im_Qv\cdot x}Q_v(x),\,\,\,\,
H=e^{-im_Qv\cdot x}H_v,\,\,\,\,
\bar{H}=e^{im_Qv\cdot x}\bar{H}_v.
\end{equation}
Note that $v$ is the heavy-quark velocity. Here, we choose $v=(+1,0,0,0)$ for definiteness, representing the heavy quark at rest~\cite{Georgi:1990um}. It is an easy task to redefine $\mathcal{L}^\mathrm{HH}_\mathrm{eff}$ in Eq.~(\ref{eq:EFL1}) using the Dirac equation for the heavy-quark field in Eq.~(\ref{eq:HMFD}), resulting in
\begin{equation}
\label{eq:HHL}
\mathcal{L}^{\mathrm{HH}}_\mathrm{eff}=\bar{Q}_v\left[\rlap{/}{v}(iv\cdot\partial)-M_Q\right]Q_v=\bar{Q}_v\left[(iv\cdot\partial)-M_Q\right]Q_v.
\end{equation}
In the second step of Eq.~(\ref{eq:HHL}), we use the velocity projection for the heavy quark, $\rlap{/}{v}Q_v=Q_v$~\cite{Georgi:1990um}. Similarly, using Eq.~(\ref{eq:HMFD}), we can rewrite $\mathcal{L}^{\mathrm{(HL,LH)}}_\mathrm{eff}$ in the following form:
\begin{equation}
\label{eq:HLLH}
\mathcal{L}^{\mathrm{HL}}_\mathrm{eff}=-\frac{1}{F_H}
\bar{Q}_v\sqrt{M_Q}
\left[\frac{1+\rlap{/}{v}}{2}i\gamma_5H_v \right]
\sqrt{M_q}q,\,\,\,\,
\mathcal{L}^{\mathrm{LH}}_\mathrm{eff}=-
\frac{1}{F_H}\bar{q}\sqrt{M_q}
\left[(i\bar{H}_v)\frac{1+\rlap{/}{v}}{2}\gamma_5 \right]
\sqrt{M_Q}Q_v.
\end{equation}
Making use of a generic functional integral technique for the Grassmann variables given for the two Grassmann variables $q$ and $Q_v$~\cite{Nam:2011ak}, finally, we can arrive at an EChA for the heavy-light quark systems from the effective Lagrangian density, given in Eq.~(\ref{eq:EFL1}):
\begin{eqnarray}
\label{eq:EFA3}
&&\mathcal{S}^\mathrm{LL+HL+LH}_\mathrm{eff}=
\cr
&&-i\mathrm{Sp}\ln
\Bigg[i\rlap{/}{\partial}-\bar{M}_q-\frac{1}{F_L}\sqrt{M_q}(i\gamma_5L)
\sqrt{M_q}-\left(\frac{1}{F_H}\sqrt{M_Q}H\sqrt{M_q}
\right) (iv\cdot\partial-M_Q)^{-1}
\left(\frac{1}{F_H}\sqrt{M_q}\bar{H}\sqrt{M_Q}\right)\Bigg].
\end{eqnarray}
It is worth noting that the effective action in Eq.~(\ref{eq:EFA3}) is in principle equivalent to the first term of Eq.~(36) in Ref.~\cite{Ebert:1994tv}, except for the momentum dependent quark-PS meson coupling strengths. 

As a next step, we estimate $M_{Q,0}$ in Eq.~(\ref{eq:MMMH}) from a simple phenomenological analysis. In this consideration, the heavy PS-meson mass can be understood as
\begin{equation}
\label{eq:HMMM}
M_H\approx
\left[m_q+M_{q,0} \right]_\mathrm{L}+\left[m_Q+M_{Q,0} \right]_\mathrm{H},
\end{equation}
where we have ignored the binding energy for the meson. From the experimental data for $D$ and $B$ mesons, we can write
\begin{eqnarray}
\label{eq:ES}
M_D&=&1869.57\,\mathrm{MeV}\approx(m_c+m_q+M_{q,0}+M_{Q,0})=1625.0\,\mathrm{MeV}+M_{Q,0}\to M_{Q,0}\approx 244.57\,\mathrm{MeV}
\cr
M_{B}&=&5279.17\,\mathrm{MeV}\approx(m_b+m_q+M_{q,0}+M_{Q,0})=5025.0\,\mathrm{MeV}+M_{Q,0}\to M_{Q,0}\approx 254.17\,\mathrm{MeV}.
\end{eqnarray}
The numerical inputs, which are taken from Ref.~\cite{Nakamura:2010zzi}, are summarized in Table~\ref{TABLE1}. As shown in Eq.~(\ref{eq:ES}), it is necessary to add the {\it effective} heavy-quark mass to reproduce the heavy-meson mass appropriately. 

Here are some explanations on this additional heavy-quark mass: In the instanton model for example, the light current quarks obtain their momentum-dependent effective masses via the nontrivial interactions with the instanton ensemble, which represents the nonperturbative QCD vacuum~\cite{Schafer:1996wv,Diakonov:1983hh,Diakonov:2002fq}. This mechanism is also equivalent to SBCS, resulting in the nonzero values for the chiral condensate $\langle\bar{q}q\rangle$ and various low-energy constants, such as $F_\Phi$. Hence, it is reasonable to assume a similar mechanism for the additional heavy-quark mass $M_{Q,0}$ in Eq.~(\ref{eq:ES}), i.e. $M_{Q,0}$ is considered to be originated from the highly nontrivial interactions between the heavy quark with $m_Q$ and the nonperturbative QCD vacuum. To estimate $M_{Q,0}$, we have chosen $M_{q,0}\approx350$ MeV as in  Section II. If we consider the binding energy for the mesons, the estimated value $M_{Q,0}=(240\sim250)$ MeV must be the lower bound of its real one. From these observations and previous discussions, we summarize our educated assumptions from a phenomenological point of view:
\begin{itemize}
\item  The finite effective heavy-quark mass, $M_{Q,0}$ is generated from the similar mechanism with that for the light quark-instanton interaction, representing the nontrivial QCD vacuum effects. For the numerical calculations, $M_{Q,0}$ can be estimated by Eq.~(\ref{eq:HMMM}).
\item If the instanton ensemble is not affected much by the heavy sources such as the heavy quarks, one uses $\Lambda_Q\approx600$ MeV again for Eq.~(\ref{eq:MMMH}) like that for the light quarks. However, we will see that the value of $\Lambda_Q$ should be changed to reproduce the experimental data for $f_\Phi$ appropriately. The difference between $\Lambda_q$ and $\Lambda_Q$ may signal the broken flavor SU(5) symmetry.
 \end{itemize}
Thus, we will use the effective heavy-quark mass in the momentum space as in Eq.~(\ref{eq:MMMH}), similar to the light-quark case. 

The heavy-meson weak-decay constants is defined as follows~\cite{Ebert:1994tv}:
\begin{equation}
\label{eq:FPIMAT}
\langle0|\bar{q}(x)\gamma_\mu(1-\gamma_5)Q_v(x)|H(p)\rangle
=ip_\mu f_H,
\end{equation}
where $p$ stands for the on-mass shell momentum of $H$ with the velocity $v$. In order to evaluate the matrix element in Eq.~(\ref{eq:FPIMAT}), we employ the external-field method as done in Ref.~\cite{Nam:2011ak}. After a straightforward functional manipulation, we obtain the following formula for the heavy-meson weak-decay constant:
\begin{eqnarray}
\label{eq:FEX}
f^2_H&=&\frac{2N_c\,\mathcal{N}_q}{\Delta M_H}\int\frac{\bm{k}^2d\bm{k}}{\pi^2}
\frac{\sqrt{M_{q,0}M_{Q,0}}(\Delta M_H-M_{Q,0})
[\bm{k}^2+(M_{Q,0}-\Delta M_H)^2+2\Lambda^2_q]^4}
{[\bm{k}^2+(M_{Q,0}-\Delta M_H)^2+2\Lambda^2_q]^4[\bm{k}^2 +(M_{Q,0}-\Delta M_H)^2]+(2\Lambda^2_q)^4\bar{M}^2_{q,0}}.
\cr
&\times&\left[\frac{2\Lambda^2_q}
{[\bm{k}^2+(M_{Q,0}-\Delta M_H)^2+2\Lambda^2_q]}
 \right]\left[\frac{2\Lambda^2_Q}{(\bm{k}^2+2\Lambda^2_Q)}
 \right],
\end{eqnarray}
where, from the on-mass shell condition for the meson $H$ at $v$~\cite{Ebert:1994tv},  one is lead to $v\cdot p\approx\Delta M_{H}\equiv M_{H}-m_{Q}$. 
\section{Numerical results}
\begin{figure}[t]
\begin{tabular}{cc}
\includegraphics[width=8.5cm]{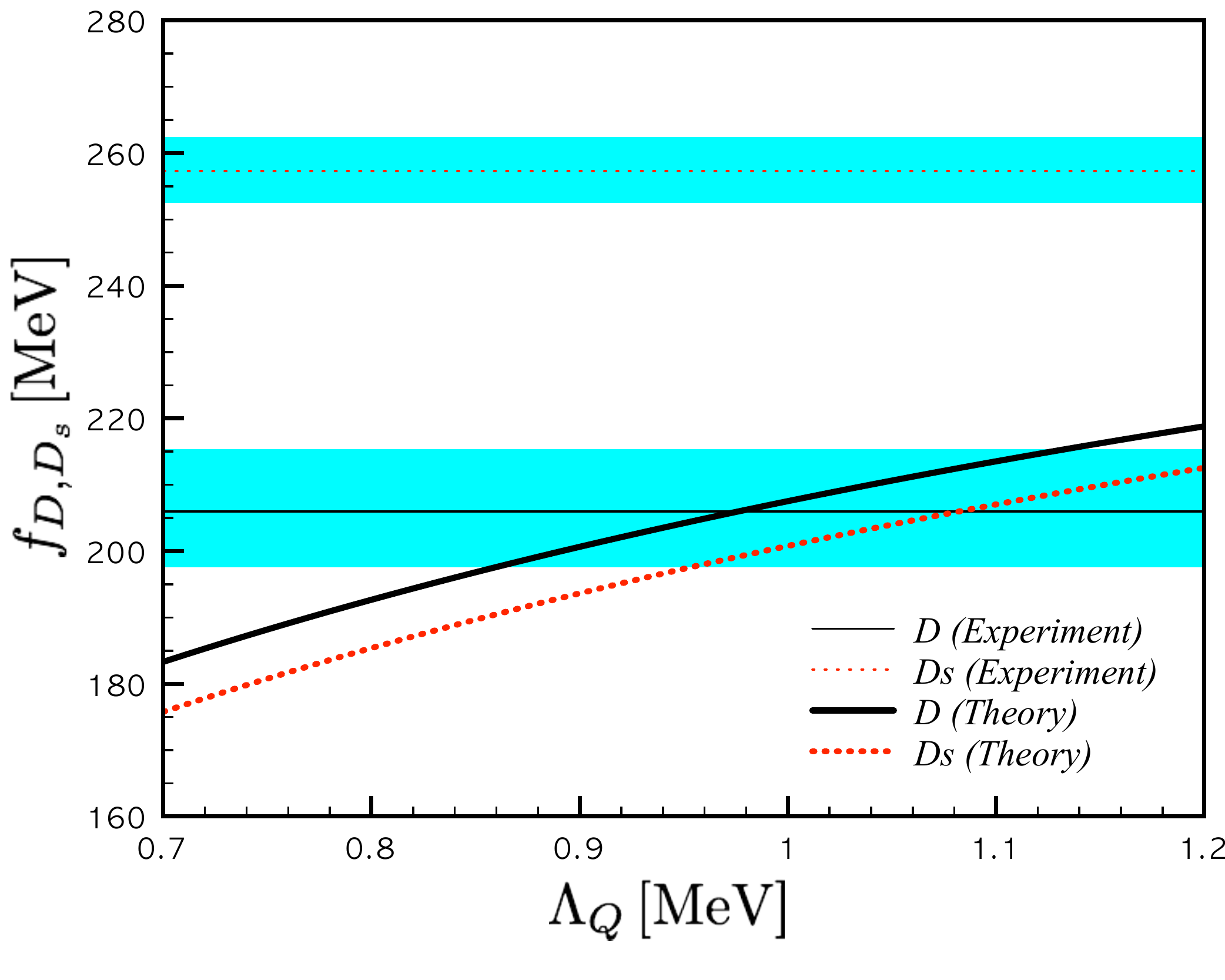}
\includegraphics[width=8.5cm]{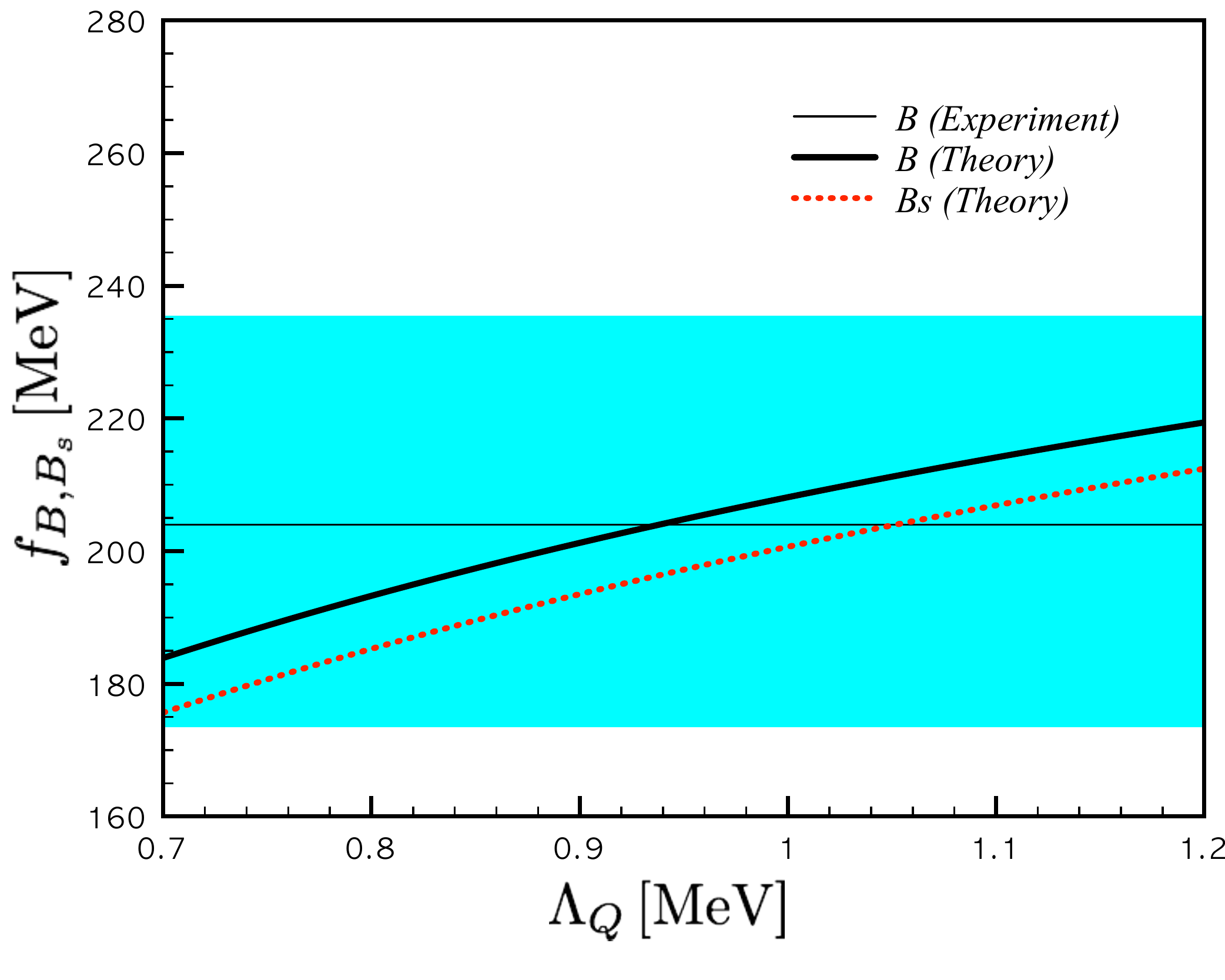}
\end{tabular}
\caption{(Color online) $f_{D,D_s}$ (left) and $f_{B,B_s}$ (right) as functions of $\Lambda_Q$ at $\Lambda_q=600$ MeV for $\mathcal{N}_{u,d,s}=1$ in Eq.~(\ref{eq:FEX}). The numerical results for the non-strange and strange mesons are shown with the solid and dot lines, respectively. The experimental data are taken from Refs.~\cite{Rosner:2008yu,Rosner:2010ak}. Note that the experimental errors are given with the shaded areas.}       
\label{FIG1}
\end{figure}

In this Section, we present the numerical results and related discussions. First, we show them for $f_{D,D_s}$ (left) and $f_{B,B_s}$ (right) for $\mathcal{N}_{q}=1$ in Eq.~(\ref{eq:FEX}) in Figure~\ref{FIG1} as functions of $\Lambda_Q$ at $\Lambda_q=600$ MeV. The curves for the strange and non-strange mesons are drawn in the solid and dot lines, respectively. Horizontal lines denote the experimental data, while the errors are represented by the shaded areas. The experimental data are taken from Refs.~\cite{Rosner:2008yu,Rosner:2010ak}. As shown in Figure~\ref{FIG1}, for the non-strange mesons, i.e. $f_{D,B}$, we see that the data are well reproduced when $\Lambda_Q\approx1$ GeV: $f_{D,B}=(207.54,\,208.13)$ MeV, where the experimental data provide $f_{D,B}=(206\pm8.9\,204\pm31)$ MeV~\cite{Rosner:2008yu,Rosner:2010ak}. Although one may choose $\mathcal{N}_{u,d}=1.17$ as done for $f_\pi$ in Section II and modify the $\Lambda_Q$ value  from $1$ GeV, we assume that the effects from $\delta\mathcal{N}_{u,d}$, which amounts the deviation from unity, i.e. $\mathcal{N}_{u,d}\equiv1+\delta\mathcal{N}_{u,d}$, are absorbed into $\Lambda_Q$ for brevity. 

Physically, this larger renormalization scale for the heavy quark $\Lambda_Q\approx1$ GeV can be understood intuitively as follows: As for the light meson, consisting of the two light quarks, the typical scale has been determined in general by $\Lambda_q\approx600$ MeV as in many instanton-motivated models like the present work. However, taking into account the heavier degree of freedom for the heavy mesons, including the heavy quarks $m_Q\sim$ a few GeV, one must introduce larger scale in comparison to the light mesons. Hence, this effect can be considered effectively as well as phenomenologically in the larger renormalization scale to reproduce the data. 

\begin{figure}[t]
\begin{tabular}{cc}
\includegraphics[width=8.5cm]{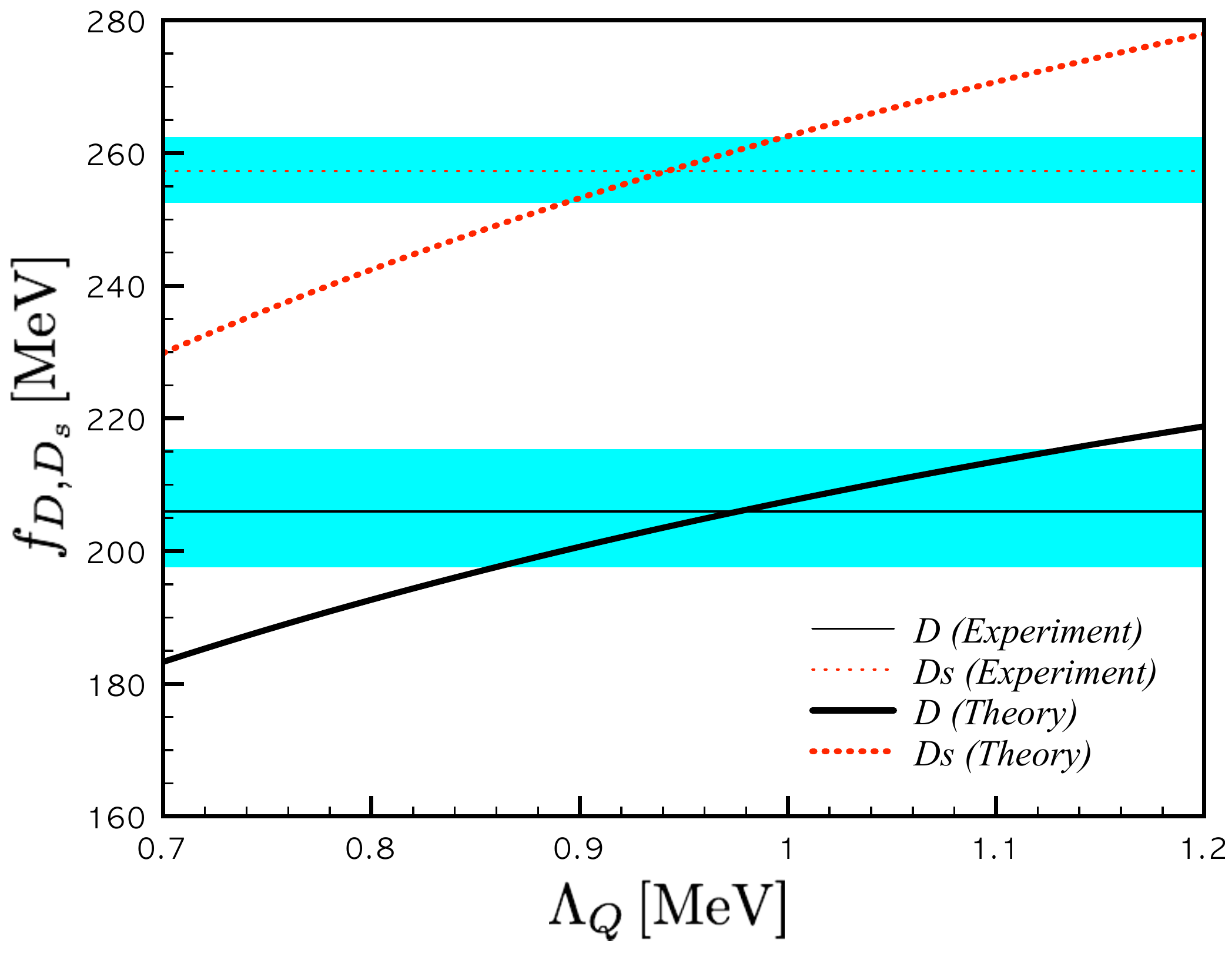}
\includegraphics[width=8.5cm]{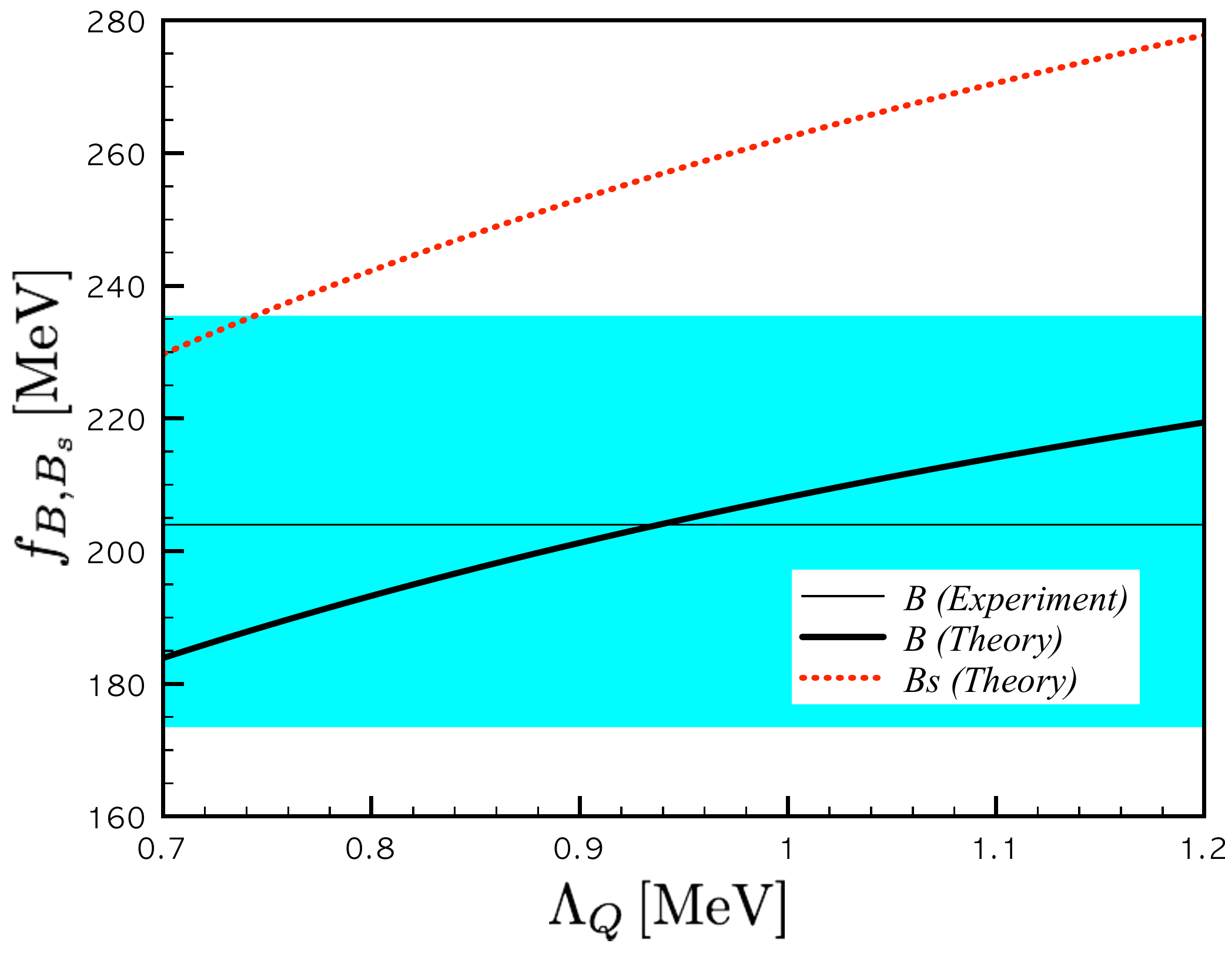}
\end{tabular}
\caption{(Color online) $f_{D,D_s}$ (left) and $f_{B,B_s}$ (right) as functions of $\Lambda_Q$ at $\Lambda_q=600$ MeV for $\mathcal{N}_{u,d}=1$ and $\mathcal{N}_s=1.71$ in Eq.~(\ref{eq:FEX}). The numerical results for the non-strange and strange mesons are shown with the solid and dot lines, respectively. The experimental data are taken from Refs.~\cite{Rosner:2008yu,Rosner:2010ak}. Note that the experimental errors are given with the shaded areas.}       
\label{FIG2}
\end{figure}

Taking this strategy, we fix $\Lambda_Q=1$ GeV for every heavy-quark flavors from now on. The numerical results for $f_{D_s}$ turn out to be much smaller than the experimental value, $f_{D_s}=200.78$ MeV with $\Lambda_Q=1$ GeV, whereas we do not have experimental data for $f_{B_s}$. These underestimations for the strange mesons can be understood by the same problem for $f_K$ with the explicit flavor SU(3) symmetry breaking as discussed in Section II. Hence, if we chose $\mathcal{N}_s=1.71$, which was introduced to reproduce the empirical value for $f_K$, the numerical results for $f_{D_s,B_s}$ are drastically improved as shown in Figure~\ref{FIG2}. Numerically, we have $f_{D_s,B_s}=(262.56,\,262.39)$ MeV to be compared with the experimental data $f_{D_s}=(257.3\pm5.3)$ MeV. From these observations, the correction factor $\mathcal{N}_s$ for the strangeness in the present model is crucial, and we have a tendency approximately from the present model:
\begin{equation}
\label{eq:TEND}
(f_D\approx f_B\approx210\,\mathrm{MeV})\,\,\,\,<
\,\,\,\,(f_{D_s}\approx f_{B_s}\approx260\,\mathrm{MeV}).
\end{equation}
Our final numerical results for $f_{H,H_s}$ and their ratio between them are listed in Table~\ref{TABLE2}. 
\begin{table}[h]
\begin{tabular}{c||c|c|c|c|c|c}
&$f_D$&$f_B$&$f_{D_s}$&$f_{B_s}$&$f_{D_s}/f_D$&$f_{B_s}/f_B$\\
\hline
\hline
Present&$207.54$&$208.13$&$262.56$&$262.39$&$1.27$&$1.26$\\
\hline
Experiment&$206\pm8.9$&$204\pm31$&$257.3\pm5.3$
&$\hspace{0.5cm}\cdots\hspace{0.5cm}$&$1.27$&$\cdots$\\
\end{tabular}
\caption{Numerical results for $f_{D,D_s,B,B_s}$ [MeV] and their ratios, computed for $\Lambda_{q,Q}=(600,1000)$ MeV, $M_{q,0}=350$ MeV, and $\mathcal{N}_{u,d,s}=(1,1,1.71)$. The experimental data are taken from Refs.~\cite{Rosner:2008yu,Rosner:2010ak}.}
\label{TABLE2}
\end{table}

\begin{figure}[t]
\begin{tabular}{cc}
\includegraphics[width=8.5cm]{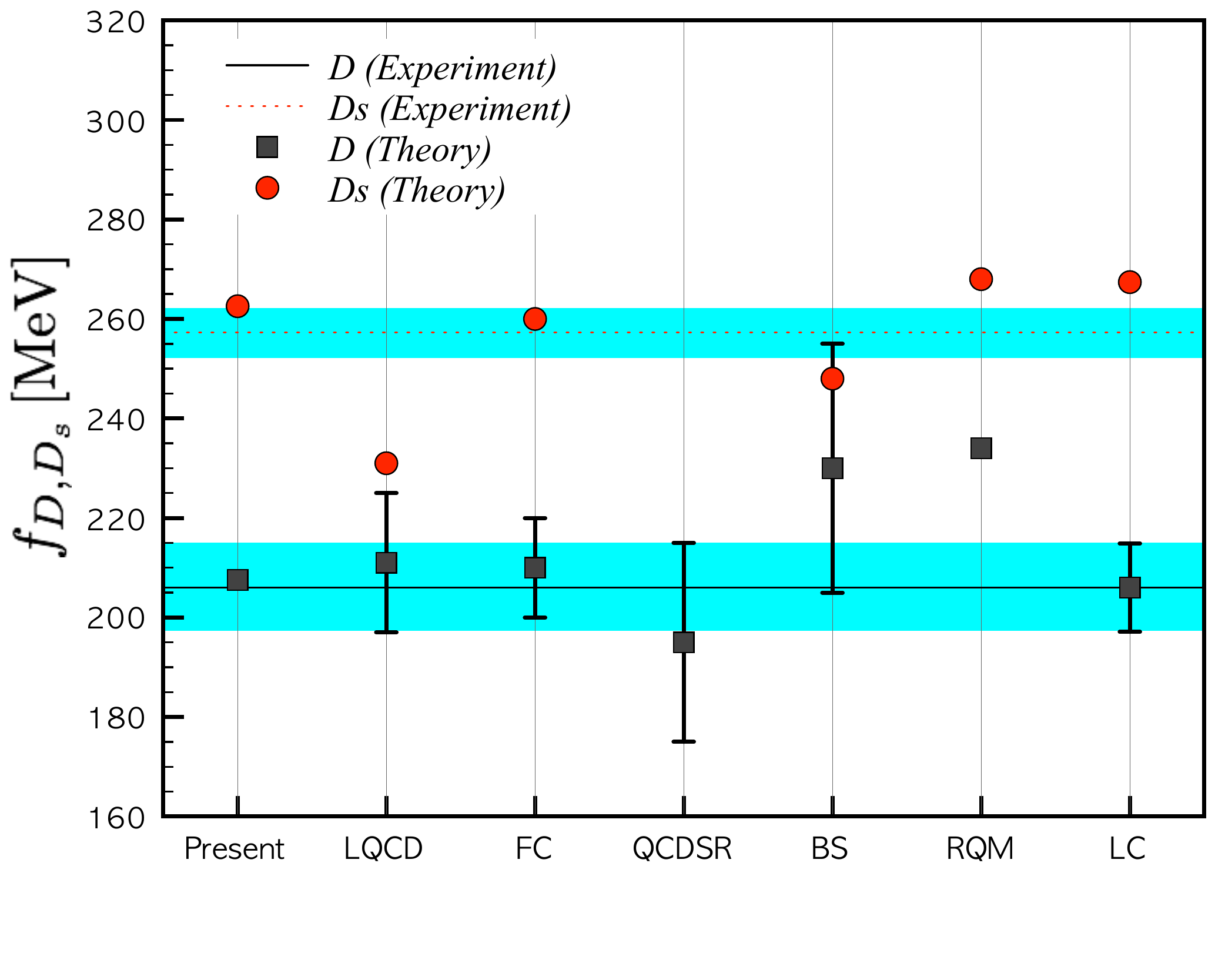}
\includegraphics[width=8.5cm]{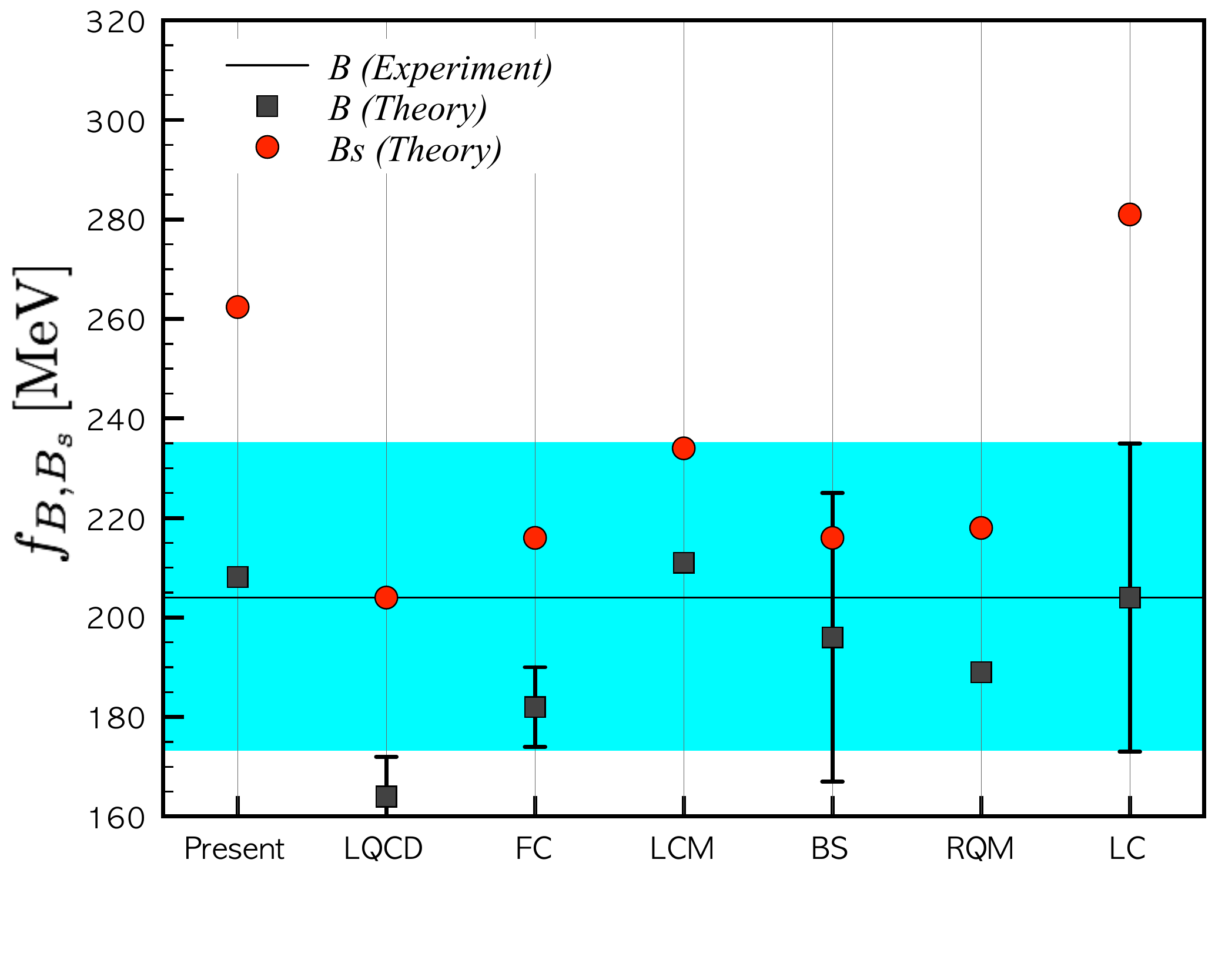}
\end{tabular}
\caption{Theoretical results for $f_{D,D_s,B,B_s}$ from the present calculations, clover-improved quenched LQCD~\cite{Becirevic:1998ua},  field-correlator method (FC)~\cite{Badalian:2007km},  QCD sum rule~\cite{Penin:2001ux}, light-front quark model (LQM)~\cite{Choi:2007se}, Bethe-Salpeter method (BS)~\cite{Cvetic:2004qg,Wang:2005qx}, relativistic quark model (RQM)~\cite{Ebert:2006hj}, and light-cone wave function (LC)~\cite{Hwang:2010hw}.}       
\label{FIG3}
\end{figure}

In Figure~\ref{FIG3}, we compare our numerical results with other theoretical estimations for $f_{D,D_s}$ (left) and $f_{B,B_s}$ (right). Other theoretical results are taken from the clover-improved quenched LQCD~\cite{Becirevic:1998ua},  field-correlator method (FC)~\cite{Badalian:2007km},  QCD sum rule (QCDSR)~\cite{Penin:2001ux}, light-front quark model (LQM)~\cite{Choi:2007se}, Bethe-Salpeter method (BS)~\cite{Cvetic:2004qg,Wang:2005qx}, relativistic quark model (RQM)~\cite{Ebert:2006hj}, and light-cone wave function (LC)~\cite{Hwang:2010hw}. As for $f_{D}$, all the theoretical results are qualitatively compatible with the data, while that from BS~\cite{Cvetic:2004qg,Wang:2005qx} overestimates considerably. In contrast, LQCD~\cite{Becirevic:1998ua} underestimates the data for $f_{D_s}$, depending on the simulation schemes. As shown in the right panel of Figure~\ref{FIG3}, the present result for $f_B$ are in good agreement with the data which contain huge error bars. All the theoretical results are still comparable with the experimental center value for $f_B=204$. The situation changes, however, drastically for $f_{B_s}$. Note that the present result and that from LC~\cite{Hwang:2010hw} give $f_{B_s}\gtrsim260$ MeV, while others locate below about $240$ MeV. Hence, the experimental measurements for $f_{B_s}$ can be a good place to test theoretical reliability of each model as shown here. Numerical values in comparison with other theoretical calculations are summarized in Table~\ref{TABLE3}.
\begin{table}[h]
\begin{tabular}{c||c|c|c|c|c|c|c|c|c|c|c}
&Present&LC~\cite{Hwang:2010hw}&LQM~\cite{Choi:2007se}&FC~\cite{Badalian:2007km}&BS~\cite{Cvetic:2004qg,Wang:2005qx}&RQM~\cite{Ebert:2006hj}&LQCD~\cite{Becirevic:1998ua}&LQCD~\cite{ElKhadra:1997hq}&LQCD~\cite{ElKhadra:1997hq}&LQCD~\cite{Aoki:1997uza}&LQCD~\cite{Blossier:2009bx}\\
\hline
$f_D$&$207.53$&$(206\pm8.9)$&$211$&$210\pm10$&$230\pm25$&$234$&$211^{+0}_{-12}\pm14$&$194^{+14}_{-10}\pm10$
&$218.9\pm11.3$&$192\pm30$&$197\pm9$\\
$f_{D_s}$&$262.56$&$267.4\pm17.9$&$248$&$260\pm10$&$248\pm27$&$268$&$231^{+6}_{-0}\pm12$&$213^{+14}_{-11}\pm11$&$260.1\pm10.8$&$214\pm33$&$244\pm8$\\
$f_B$&$208.13$&$(204\pm31)$&$189$&$182\pm8$&$196\pm29$&$189$&$179^{+26}_{-9}\pm18$&$164^{+14}_{-11}\pm8$&$196.9\pm8.9$&$171\pm29$&$\cdots$\\
$f_{B_s}$&$262.39$&$281\pm54$&$234$&$216\pm8$&$216\pm32$&$218$&$204^{+28}_{-0}\pm16$&$185^{+13}_{-8}\pm9$&$242\pm9.5$&$193\pm32$&$\cdots$\\
\end{tabular}
\caption{Theoretical results for $f_{D,D_s,B,B_s}$ from the present calculations, light-cone wave function (LC)~\cite{Hwang:2010hw}, light-front quark model (LQM)~\cite{Choi:2007se}, field-correlator method (FC)~\cite{Badalian:2007km}, Bethe-Salpeter method (BS)~\cite{Cvetic:2004qg,Wang:2005qx}, relativistic quark model (RQM)~\cite{Ebert:2006hj}, clover-improved quenched LQCD~\cite{Becirevic:1998ua}, $\mathcal{O}(a)$-improved light-quark action~\cite{ElKhadra:1997hq}, $(2+1)$-flavor asqdat action~\cite{ElKhadra:1997hq}, full one-loop correlator~\cite{Aoki:1997uza}, and Symanzik gauge action~\cite{Blossier:2009bx}. }
\label{TABLE3}
\end{table}

In what follows, we extend our discussions to very rough estimations for the heavy-meson nonrelativistic wave function and CKM matrix elements, based on the numerical results obtained. In the nonrelativistic quark models, the heavy PS-meson decay constant is given by the following formula~\cite{VanRoyen:1967nq}:
\begin{equation}
\label{eq:FWF}
f^2_H=\frac{12}{M_H}|\Psi_H(0)|^2,
\end{equation}
where $\Psi_H(0)$ is the wave function for the heavy meson at zero separation between the heavy and light quarks inside. Using the computed values for $f_H$ as shown in Table~\ref{TABLE2}, we obtain
\begin{equation}
\label{eq:PSI}
|\Psi_{D,D_s,B,B_s}(0)|=(0.08,0.11,0.14,0.18),
\end{equation}
which result in the following ratios, 
\begin{equation}
\label{eq:PSI2}
\frac{|\Psi_{D}(0)|}{|\Psi_{D_s}(0)|}=0.73,\,\,\,\,
\frac{|\Psi_{B}(0)|}{|\Psi_{B_s}(0)|}=0.78,,\,\,\,\,
\frac{|\Psi_{D}(0)|}{|\Psi_{B}(0)|}=0.57,\,\,\,\,
\frac{|\Psi_{D_s}(0)|}{|\Psi_{B_s}(0)|}=0.61.
\end{equation}
Note that the ratio between $D(D_s)$ and $B(B_s)$ is considerably smaller than unity, being different from the rough assumption made in Ref.~\cite{Rosner:1990xx}, whereas the estimations for the ratio between $D(B)$ and $D_s(B_s)$ are relatively similar to each other. These numerical results for the ratios will be informative to construct and estimate the heavy-meson wave functions in the nonrelativistic quark models.

Now, we want to apply the numerical results for $f_{D,D_s,B,B_s}$ to some phenomenological problems. As known well, the PS-meson leptonic decay is a good subject to study the weak interactions corresponding to the CKM matrix elements. Here, $f_H$ plays a critical role. The partial decay width for $P\to\ell\nu$ reads:
\begin{equation}
\label{eq:GAMMALEP}
\Gamma(P\to\ell\nu)=\frac{G^2_F}{8\pi}f^2_P\, M_P\, M^2_\ell\,
\left(1-\frac{M^2_\ell}{M_P} \right)^2|V_{fg}|^2,
\end{equation}
where $G_F$, $P$, $f_P$, $M_P$, and $M_\ell$ denote the Fermi constant, generic PS meson, PS-meson weak-decay constant, PS-meson mass, and lepton mass, respectively. $V_{fg}$ stands for the CKM matrix element for the quark flavors $f$ and $g$. As for $D^+\to\mu^+\nu$, from the CLEO experiment, the branching ratio $\mathcal{B}(D^+\to\mu^+\nu)$ was given by $(3.82\pm0.32\pm0.09)\times10^{-4}$~\cite{Nakamura:2010zzi}. Note that $\mathcal{B}$ is defined by $\tau_P\Gamma(P\to\ell\nu)/\hbar$, in which $\tau_P$ denotes the life time of the PS meson. Using the numerical result for $f_D$ given in Table~\ref{TABLE2}, we have the CKM matrix element $|V_{cd}|=0.224$, whereas it becomes $0.225$ for the experimental $f_D$ value. Similarly, as for $D^+_s$ and $B^+$ from the present (experiment) results, we have $|V_{cs}|=0.968\,(0.975)$ and $|V_{ub}|<5.395\,(5.501)\times10^{-3}\,(90\%\,\mathrm{CL})$, using $\mathcal{B}(D^+_s\to\mu^+\nu)=5.8\times10^{-3}$ and $\mathcal{B}(B^+\to\mu^+\nu)<1.0\times10^{-6}$~\cite{Nakamura:2010zzi}. Note that the PDG fits provide $|V_{cd}|=0.2252\pm0.0007$, $|V_{cs}|=0.97345^{+0.00015}_{-0.00015}$, and $|V_{ub}|=(3.47^{+0.16}_{-0.12})\times10^{-3}$. As understood, our numerical results are in good agreement with the PDF fits. All the numerics for the CKM matrix elements are listed in Table~\ref{TABLE4}.

The mesonic decay of $\bar{B}^0$ can be computed by the following formula~\cite{Rosner:1990xx}:
\begin{equation}
\label{eq:MESONIC}
\Gamma(\bar{B}^0\to D^+P^-)=\frac{G^2_F}{128\pi}
|V_{cb}|^2|V_{fg}|^2\,M^3_B\,f^2_P\,|\mathcal{F}(w^2_P)|^2\,
\sqrt{\frac{\lambda\left(1, \mathcal{A}, \mathcal{B}\right)}{\mathcal{A}}}
\{(1-\sqrt{\mathcal{A}})[(1+\sqrt{\mathcal{A}})-\mathcal{B}]\}^2,
\end{equation}
where we have used the notations $\mathcal{A}\equiv M^2_D/M^2_B$ and $\mathcal{B}\equiv M^2_P/M^2_B$, and $\lambda(x,y,z)$ indicates the following kinematic function, i.e.
\begin{equation}
\label{eq:KEOLLEN}
\lambda(x,y,z)=[x^2-(y+z)^2][x^2-(y-z)^2].
\end{equation}
The universal form factor $\mathcal{F}_P$ for the PS meson can be parameterized as follows:
\begin{equation}
\label{eq:FFF}
\mathcal{F}_P(w^2_P)=\left[\frac{\alpha_s(m^2_b)}{\alpha_s(m^2_c)} \right]^{-\frac{6}{32-2N_f}}
\frac{1}{1-w^2_P/w^2_0}.
\end{equation}
Here are two dimensionless quantities: $w_P=[M^2_P-(M_B-M_D)^2]/(M_BM_D)$ and $w_0$ corresponds to the cutoff mass for the form factor parameterization which can be determined by experiments and/or theories. In Ref.~\cite{Rosner:1990xx}, $w_0$ was estimated as $1.12\pm0.17$ that we will use for further discussions below. We also choose $\alpha_s(m^2_b)=0.189$ and $\alpha_s(m^2_c)=0.29$~\cite{Rosner:1990xx}.
Using Eq.~(\ref{eq:MESONIC}) for $P=(D^-,D^-_s)$, we can write a ratio as follows:
\begin{equation}
\label{eq:RATIOBBS}
\frac{\mathcal{B}(\bar{B}^0\to D^+D^-)}{\mathcal{B}(\bar{B}^0\to D^+D^-_s)}
=1.21\times\tan^2\theta_C\times \left(\frac{f_D}{f_{D_s}} \right)^2=0.76\times\tan^2\theta_C,
\end{equation}
where $\tan\theta_C\equiv|V_{us}|/|V_{ud}|$ with the Cabibbo angle $\theta_C$. Considering the experimental data for the branching ratios without mixings, $\mathcal{B}(\bar{B}^0\to D^+D^-)=(2.11\pm0.31)\times10^{-4}$ and $\mathcal{B}(\bar{B}^0\to D^+D^-_{s})=(7.2\pm0.8)\times10^{-3}$~\cite{Nakamura:2010zzi}, one obtains $\theta_C\approx11.11^\circ$. Note that, if we choose $\mathcal{F}(w^2_D)\approx\mathcal{F}(w^2_{D_s})$ in Eq.~(\ref{eq:MESONIC}) as in Ref.~\cite{Bhattacharya:2004fp}, $\theta_C$ becomes about $12.36^\circ$. These estimations are compatible with the empirical value $\theta_C\approx13^\circ$~\cite{Nakamura:2010zzi}.

The CKM matrix elements with $t$ flavor can be indirectly obtained by the following equation~\cite{Nakamura:2010zzi}:
\begin{equation}
\label{eq:DELTARATIO}
\frac{\Delta m_{B_s}}{\Delta m_{B}}=\frac{M_{B_s}}{M_{B}}\xi_B^2
\left|\frac{V_{ts}}{V_{td}}\right|^2,
\end{equation}
where $\Delta m_{B,B_s}$ stand for the mass difference between the heavy and light mass eigenstates of the $B^0$-$\bar{B}^0$ mixing in the standard model (SM), so called as the oscillation frequencies~\cite{Nakamura:2010zzi}, in which those values are given as $\Delta m_B=(0.507\pm0.005)\,\mathrm{ps}^{-1}$ and $\Delta m_{B_s}=[17.77\pm0.10\,(\mathrm{stat})\pm0.07\,(\mathrm{syst})]\,\mathrm{ps}^{-1}$.  $\xi_B$ is the flavor SU(3) symmetry breaking factor, defined by
\begin{equation}
\label{eq:XIB}
\xi_B=\sqrt{\frac{B_{B_s}}{B_{B}}}\frac{f_{B_s}}{f_{B}}.
\end{equation}
Here, $B_{B,B_s}$ denote the bag parameter appearing in the SM calculations for the $B^0\to\bar{B^0}$ transition. If we set their ratios to be unity as done in Ref.~\cite{Bhattacharya:2004fp}, we have from the present numerical results $\xi_B=1.26$, which is well compatible with that given in the LQCD simulation~\cite{Gray:2005ad}, $\xi_{B,\mathrm{LQCD}}=1.21^{+0.046}_{-0.035}$. Ref.~\cite{Laiho:2009eu} gives $\xi_B=1.243\pm0.021\pm0.021$ via a global fit using the LQCD data. Plugging the empirical values for $\Delta m_{B,B_s}$ and $M_{B,B_s}$ into Eq.~(\ref{eq:DELTARATIO}) and using the numerical result for $\xi_B$, we have $|V_{td}/V_{ts}|=0.215$, which is similar to the PDG fit, $\sim0.214$. Note that a LQCD simulation~\cite{Laiho:2009eu} provides $|V_{td}/V_{ts}|=0.211\pm0.001\pm0.005$~\cite{Nakamura:2010zzi}. The numerical results are summarized in Table~\ref{TABLE4}.
\begin{table}[h]
\begin{tabular}{c||c|c|c|c}
&$|V_{cd}|$&$|V_{cs}|$&$|V_{ub}|\times10^{-3}$&$|V_{td}/V_{ts}|$\\
\hline
\hline
Present&$0.224$&$0.968$&$<5.395$&0.215\\
\hline
PDG fit&$0.2252\pm0.0007$&$0.97345^{+0.00015}_{-0.00015}$&$3.47^{+0.16}_{-0.12}$&$0.211\pm0.001\pm0.005$
(LQCD~\cite{Laiho:2009eu,Nakamura:2010zzi})\\
\end{tabular}
\caption{CKM matrix elements including the heavy-quark flavors, $|V_{cd}|$, $|V_{cs}|$, and $|V_{ub}|$.}
\label{TABLE4}
\end{table}

\section{Summary and perspectives}
In the present work, we have investigated the weak-decay constants for the heavy PS-mesons consisting of the $(u,d,s,c,b)$ flavors, i.e. $f_{D,D_s,B,B_s}$. To this end, we employ flavor SU(5)  ExNLChQM which was developed being motivated by HQEFT and NLChQM. We chose the cutoff masses for the light and heavy quarks differently as $\Lambda_q=600$ MeV and $\Lambda_Q=1000$ MeV, respectively, considering that the nontrivial QCD vacuum effects are finite even for the heavy quark inside the meson. In what follows, we list important observations in the present work:
\begin{itemize}
\item We employ a phenomenological correction factor for the inclusion of the strange quark, compensating the $1/N_c$ corrections. By doing that, we obtain $f_{D,D_s,B,B_s}=(207.53,\,262.56,\,208.13,\,262.39)$ MeV, which are qualitatively compatible with available experimental and theoretical values. Thus, this correction factor turns out to be crucial to reproduce the data.
\item In other words, this qualitatively good agreements with the data indicate considerable contributions for the heavy quarks from the nontrivial QCD-vacuum effects. Moreover, it is justified that the vacuum structure due to the heavy quarks is modified in comparison to that for the light quarks by seeing that $\Lambda_q\ne\Lambda_Q$, since $\Lambda$ value denotes the average size of (anti)instantons. This tendency also signals the breakdown of the SU(5) flavor symmetry at the quark level, in addition to $m_q\ne m_Q$.
\item The nonrelativistic heavy-meson wave functions at zero separation between the light and heavy quarks, $\Psi_H(0)$, are calculated. From those numerical results, we observe that the ratio between the wave functions for the strange and non-strange heavy mesons is less than unity. This observation is quite different from simple consideration in quark models, i.e. $|\Psi_{H}(0)|/|\Psi_{H_s}(0)|=1$.   
\item The CKM matrix elements are estimated by using the various leptonic decays of the heavy PS mesons, resulting in that $(|V_{cd}|,|V_{cs}|,|V_{ub}|)=(0.224,0.968,<5.395\times10^{-3})$. Taking into account the mass difference between the heavy and light mixing of $B^0$-$\bar{B}^0$ states, although there are uncertainties arising from the bag parameters in SM, we have $|V_{td}|/|V_{ts}|=0.215$. Again, these values are well comparable with their empirical and LQCD simulation data. 
\item Accounting for the hadronic decays for the heavy meson and their ratios, we compute the flavor SU(3) symmetry breaking factor as $\xi_B=1.26$. The Cabibbo angle can be also estimated using this information, resulting in $\theta_C=12.36^\circ$, depending on the choice of the PS-meson form factor. We note that these values are consistent with its empirical data and LQCD results.
\end{itemize}

Taking into account all the results summarized above, we can conclude that the present theoretical framework, ExNLChQM is a qualitatively successful model to study the heavy PS meson, which consist of the heavy and light quarks. As for the future improvements of the present form of ExNLChQM, 1) the nonlocal contributions for conserving the axial-vector current can be included to compute the weak-decay constants. Furthermore, 2) the $1/N_c$ corrections can be included in the present framework in principle via standard functional treatments, instead of using the phenomenological correction factor, although this task must quite challenging. Nonetheless for these possible improvements, from a phenomenological point of view, the present theoretical framework is still a considerably useful tool to investigate various physical quantities for the heavy-light quark systems, such as the Isgur-Wise function, heavy-light meson coupling constants, heavy-meson effective Lagrangians, and so on. Related works are under progress and appear elsewhere.

\section*{acknowledgment}
The author is grateful to P.~Ko (KIAS) and C.~Yu (KIAS) for fruitful discussions and comments.

\section*{Appendix}
The SU(5) PS-meson fields are defined as follows:
\begin{equation}
\label{eq:U5F}
\mathrm{U}_5=\sqrt{2}\left(
\begin{array}{ccccc}
\frac{\pi^0}{\sqrt{2}}+\frac{\eta}{\sqrt{6}}+\frac{\eta_c}{\sqrt{12}}+\frac{\eta_b}{\sqrt{20}}&\pi^+&K^+&\bar{D}^0&B^+\\
\pi^-&-\frac{\pi^0}{\sqrt{2}}+\frac{\eta}{\sqrt{6}}+\frac{\eta_c}{\sqrt{12}}+\frac{\eta_b}{\sqrt{20}}&K^0&D^-&B^0\\
K^-&\bar{K}^0&-\frac{2\eta}{\sqrt{6}}+\frac{\eta_c}{\sqrt{12}}+\frac{\eta_b}{\sqrt{20}}&D^-_s&B^0_s\\
D^0&D^+&D^+_s&-\frac{3\eta_c}{\sqrt{12}}+\frac{\eta_b}{\sqrt{20}}&B^+_c\\
B^-&\bar{B}^0&\bar{B}^0_s&B^-_c&-\frac{4\eta_b}{\sqrt{20}}\\
\end{array}
\right).
\end{equation}


\end{document}